\documentclass[twocolumn,english, superscriptaddress,10pt]{revtex4-2}

\usepackage[T1]{fontenc}
\usepackage[utf8]{inputenc}

\usepackage[colorlinks=true,linkcolor=blue,urlcolor=blue,citecolor=blue,breaklinks=true]{hyperref}
\usepackage{xurl}

\usepackage{amsmath}

\usepackage{graphicx}

\usepackage{makecell}
\usepackage{booktabs}

\begin{document}
\title{The stochastic nature of power-grid frequency in South Africa}

\author{Leonardo~Rydin~Gorj\~ao}
\affiliation{Department of Computer Science, OsloMet -- Oslo Metropolitan University, N-0130 Oslo, Norway}
\affiliation{Faculty of Science and Technology, Norwegian University of Life Sciences, 1432 Ås, Norway}

\author{Jacques~Maritz}
\affiliation{Department of Engineering Sciences, University of the Free State, Bloemfontein 9301, South Africa}

\begin{abstract}
In this work, we explore two mechanisms that explain non-Gaussian behaviour of power-grid frequency recordings in the South African grid.
We make use of a Fokker--Planck approach to power-grid frequency that yields a direct relation between common model parameters such as inertia, damping, and noise amplitude and non-parametric estimations of the same directly from power-grid frequency recordings.
We propose two explanations for the non-Gaussian leptokurtic distributions in South Africa:
The first based on multiplicative noise in power-grid frequency recordings, which we observe in South Africa;
The second based on the well-known scheduled and unscheduled load shedding and rolling blackouts that beset South Africa.
For the first we derive an analytic expression of the effects of multiplicative noise that permits the estimation of all statistical moments -- and discuss drawbacks in comparison with the data;
For the second we employ a simple numerical analysis with a modular power grid of South Africa.
Both options help understand the statistics of power-grid frequency in South Africa -- particularly the presence of heavy tails.
\end{abstract}

\maketitle

\section{Introduction}

Power-grid systems power the world.
The developed countries rely on the availability of electrical power to the extent that no less than a 24-hour service is conceivable.
Yet this is a reality only in parts of the world, so much that this is one of the key elements of the Sustainable Development Goals (SDG) --
SDG 7: ``Ensure access to affordable, reliable, sustainable and modern energy for all''~\cite{SDG2015}.
One curious example of a thus far poorly managed power system is the case of South Africa, a country still dominantly supplied via coal-fired power plants~\cite{Thopil2018}.
At least 12\% of the population does not have access to power and roughly 10\% cannot adequately afford electricity~\cite{Bohlmann2018, Lawrence2020, Ayamolowo2022}, particularly in rural areas~\cite{Meyer2021}.
Nevertheless, the power system has continuously evolved and has successfully attracted renewable-energy centred investments~\cite{Jain2017, Thopil2018}.
Yet these developments stand on a frail and deteriorating infrastructure. 
Since 2007 the maintenance and development of the South African grid has been hindered by mismanagement~\cite{Baker2019, Buraimoh2020}.
Mismanagement has manifested itself in recent years with its inability to deliver power according to the country's demand.
This lead to the implementation of `rolling blackouts' load shedding events across the country~\cite{SAloadshed}.
Load shedding has marked deleterious societal effects~\cite{Gehringer2018, Laher2019, Kobina2019}.
In 2021, the citizens and industries of South Africa were afflicted by a lack of power and periodic load shedding for over $48$ days of the year.
Additional to load shedding, the utility experiences the overloading of transformers and substations due to illegal connections, overloading, and vandalism.
The latter adds to unplanned outages (also known as non-technical losses) for parts of the network and accounts for $10\%$ of the non-technical losses~\cite{Louw2019}. 
Energy losses experienced by the utility are due to meter tampering, incorrect billing, and cable theft~\cite{Khonjelwayo2021}.
The overloading of electrical infrastructure leads to mandatory disconnection of municipalities from the distribution network.

Lack of power supply or the need for abrupt load-shedding leads to visible changes in power-grid frequency, the key signature of a power-grid system~\cite{Kundur1994, Machowski2020}.
Alternating current power grids operate at a nominal frequency -- $50$\,Hz in South Africa -- which means all conventional power-generating units contain a mechanical rotating generator tuned to this frequency.
If the power generation falls short of the power demand, frequency sags.
If the power generator surpasses demand, frequency rises.
Power grids are complex systems with inherently complex stochastic dynamics~\cite{Pagani2013, Schaefer2018}.
Power generation, both conventional but particularly renewable, is manifestly volatile~\cite{Anvari2016, Schmietendorf2017, Wolff2019, Qin2022} and non-Gaussian behaviour permeates the statistics of most power system~\cite{Schaefer2018, RydinGorjao2021a, Tyloo2022b}.
Similarly, the demand side can be equated to a complex stochastic process~\cite{Han2022b}.
Thus, we take a complex-system approach to modelling and analysing power-grid systems from a complex system perspective~\cite{Friedrich2011, Tabar2019}. 

In this work, we will examine three power-grid frequency recordings from 2021, from Continental Europe, the Nordic Grid, and the South African grid, with a focus on the latter.
We investigate these recordings and contrast their statistical and stochastic characteristics by employing a phenomenological Fokker--Planck equation of an aggregated swing-equation model with stochastic noise~\cite{Acebron1998, Acebron2000, Filatrella2008, Rodrigues2016}.
We seek an explanation for the non-Gaussian distribution of power-grid frequency observed in South Africa, wherein we offer two explanations for the heavy tails of the distributions~\cite{Pagani2013,Schaefer2018,RydinGorjao2021a}.
Firstly we identify the potential presence of multiplicative noise in the recordings and find an analytic approximation for its effect on the statistical moments of Langevin processes with multiplicative noise.
We contrast this with two other recordings of power-grid frequency in South Africa and discuss the limitations of such an approach.
Lastly, we simulate power-grid frequency trajectories in a model grid of South Africa and mimic the effect of load shedding.
Both approaches help us explain the non-Gaussian statistics of the power-grid frequency recordings of South Africa.

This work is organised as follows.
In Sec.~\ref{sec:2} we shortly introduced the historical context situating the South African power grid, the ongoing load-shedding events, and we showcase the direct effects at the University of the Free State in Bloemfontein.
In Sec.~\ref{sec:3} we introduce both a low-level dynamical model of power-grids affected by noise as well as Fokker--Planck approximation of the power-grid frequency statistics.
From this basis, we present our theoretical and numerical results.
In Sec.~\ref{sec:3} we discuss the outcomes and limitations of our data examination as extend some thoughts on future work and the necessity to examine previously overlooked power grids.

\section{Background}\label{sec:2}

\subsection{Load-shedding in South Africa}

In this work, we focus on power-grid frequency recordings from the Continental European power grid, the Nordic Grid, and the South African power grid.
It is the latter that interests us the most for this examination.
The South African power grid is undergoing an unprecedented transformation.

The power sector in South Africa is dominated by the state-owned entity ESKOM, which both imports and exports energy from surrounding states with some input from nuclear and renewable energy (photovoltaic (PV), wind and concentrated solar power (CSP)~\cite{ESKOMReport2021}.
Unfortunately, the South African power system has been experiencing an energy crisis since 2007 with electrical generation lagging the electrical demand, ultimately leading to implemented rolling blackouts in an effort to stabilise the national grid.

Due to extended periods of lack of maintenance and mismanagement resulted in an unpredictable and unreliable power system in South Africa, that during 2021 experienced load-shedding for $1169$ hours with $1775$\,GWh (with the upper limit of $2521$\,GWh) energy shed.
The latter amounts to 13.3$\%$ of all hours of 2021 being in the state of, mostly, Stage 2 load-shedding (for details see Tab.~\ref{tab:1} in Sec.~\ref{sec:num_sim})~\cite{SAloadshed}.
The power system remained coal powered during 2021, but with $18.6\%$ contributed by renewable energy (of which $6.7\%$ were variable sources)~\cite{Thopil2018}.
Concernedly, $3.2$\,TWh of energy was generated by using diesel. 

With critical low reserve margins and increasing load-shedding, some consumers are considering the implementation of self-sufficient microgrids with some renewable input.
The latter is preceded by the digitisation of the existing electrical network topology (i.e., smart grids) that requires initial capital input~\cite{Motjoadi2020}. 
South African university campuses are naturally forced to evolve according to the constraints of the national utility, the ability to swiftly react to demand reduction signals being the most critical, see Fig.~\ref{fig:1}. 
In cases of severe loss of supply, self-sufficient microgrids are utilised that synchronise with available renewable supply to ensure supply continuity on campus.
The latter typically being hybrid PV-Diesel based microgrids.

During load-shedding events (i.e. demand overpowering generation), the utility reduces load in stages of 1000\,MW.
A total of 6 stages of load-shedding are in place in South Africa at this moment, with varying duration, usually designed to stay in place for $i\times 6$ hours, with $i$ the stage.
An overview of the last three years of total load shed energy can be found below in Tab.~\ref{tab:1} (and in Ref.~\cite{SAloadshed}.).
One should note that 1) load-shedding upper limits are never actually achieved; 2) load-shedding hours are uncertain due to additional switching time; 3) additional outages exist after switching.
Unfortunately, access to data relating to all these events is hard, and thus we focus primarily on modelling uncertain load-shedding events with an equivalent proportion of hours affected vs total hours in 2021.
Therein we consider the occasional events of total blackout, which in our case is represented by the load of a region vanishing for a period of time.

\subsection{The visible consequences at the University of the Free State}

\begin{figure}[t]
    \includegraphics[width=\columnwidth]{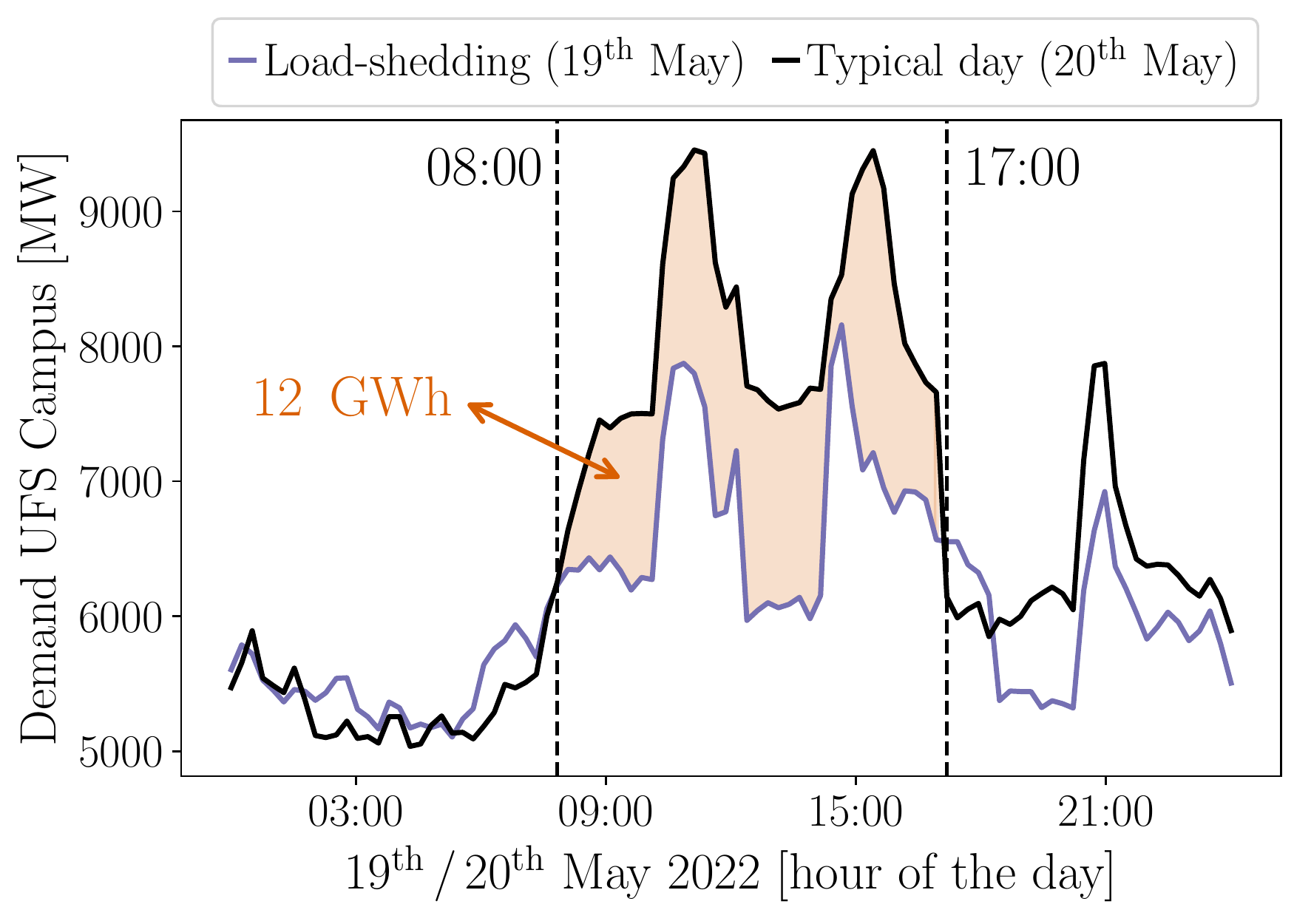}
    \caption{Demand-side management (as response to utility load-shedding) at the University of the Free State (UFS) Campus in Bloemfontein, South Africa, in comparison with a typical day.
    This is one solution among many that public and private institutions have had to implement to be able to operate uninterruptedly.
    Displayed are the 19\textsuperscript{th} and 20\textsuperscript{th} of May 2022, the first of this a day with campus demand response totalling $\sim\!\!12\,$GWh, the second a typical day without the call of demand response}\label{fig:1}
\end{figure}

The University of the Free State (UFS) Bloemfontein campus has a remarkable resource in the form of a campus smart grid.
The latter enables granular control and monitoring of intensive loads based on optimised scenarios or \textit{in-situ} demand response requests from the utility.
Dual metering on both sides of the utility feed enables monitoring of the UFS load, during instances of load-shedding, to determine the reduction needed to comply with the requirements set out by the utility. 
Maintaining high levels of control granularity ensures that the institution can act as a packetised energy manager~\cite{Espinosa2018}, hence responding to the exact requirements of the power grid operator and generating the opportunity of ancillary frequency support to the power system using the available generating resources on campus. 
Fig.~\ref{fig:1} illustrates the concept of acting as a packetised energy manager, ultimately responding to the requirements of the utility to reduce demand for a specified duration. 
Demand response to utility curtailment notices is driven by the ability to reduce demand for a negotiated time window while remaining connected to the power grid, hence frequency support is inherent, especially in the case of UFS campus power systems that are largely supported by photovoltaic renewable generators.
De-loaded photovoltaic plants can provide frequency support via the control of plant active power~\cite{Liao2015}. 
Load-shedding becomes the critical element when designing rural microgrids, with the emphasis moving from economic sense to power continuity, ultimately, converging in microgrids that couple with existing PV generation and require islanded inertia and sensitive frequency control.

\section{Statistics of power-system frequency recordings}\label{sec:3}

\subsection{The swing equation}
An operating AC power-grid system is a set of highly synchronised oscillators operating at a rotational frequency multiple of $50$\,Hz (or $60$\,Hz).
In a reduced format, we can consider a general model for power grids as coupled inertial oscillators in a network.
A synchronous machine $j$, a generator or a load, is described by its rotor-angle $\theta_j(t)$ and its angular frequency $\omega_j(t)=\dot{\theta}_i(t)$.
It obeys the equations of motion~\cite{Machowski2020, Schmietendorf2014, Sharafutdinov2018, Boettcher2022}
\begin{equation}
\begin{aligned}\label{eq:swing}
    \dot \theta_j&=\omega_j,\\
    M_j \dot \omega_j&=-D_j \omega_j + P_j^{\mathrm{m}} - P_j + \xi_j,
\end{aligned}
\end{equation}
where $M_j$ is the inertial mass of the $j$-th rotating machine, $D_j$ its damping factor, $P_j^{\mathrm{m}}$ comprises the self-generated mechanical power (or if negative it acts as a load), and $P_j$ is the exchanged power with the other oscillators in the network.
The innocently looking exchange power $P_j$ embodies the very complex structure of power-grid systems, i.e., $P_j$ is given by
\begin{equation}\label{eq:power_exchange}
    P_j = \sum_{\ell=1}^N\! E_{j} E_{\ell} \left[B_{j,\ell} \sin(\delta_j\!-\!\delta_\ell) \!+\! G_{j,\ell} \cos(\theta_j\!-\!\theta_\ell)\right],
\end{equation}
where $E_j$ is the transient voltage at $j$-th machine, which is our reduced swing equation model~\eqref{eq:swing} is static.
The parameters $G_{j,\ell}$ and $B_{j,\ell}$ denote the real and imaginary parts of the nodal admittance matrix $Y_{j,\ell}$ and encode the network structure.
The conductance $G_{j,\ell}$ comprises all the resistive terms of the network which we will not deal with, i.e., we will consider the case of a purely lossless system $G_{j,\ell}=0, \forall j,\ell$.
Conductance effects can be disregarded as they are necessarily balanced in a functional power system where the generated power is higher than the consumed one to compensate for these losses.
These models are also known as second-order/inertial Kuramoto models~\cite{Acebron2000, Filatrella2008, Schmietendorf2014, Witthaut2022}.

The network topology is fully embedded in the susceptance matrix $B_{j,\ell}$ which, apart from shunt susceptances, is equivalent to the graph Laplacian of the network~\cite{Newman2018}.
It should be noted that the swing equation~\eqref{eq:swing} or other more complex models only have a descriptive character of what happens in functioning power-grid systems.
For example, in our case, we will simulate power-grid frequency trajectories in South Africa using a very reduced model of the real-world South African power-grid system, wherein we consider a reduced number of nodes in a coupled graph of interacting generators/loads.
These coarse-grained models are illustrative at best -- yet they capture with surprising accuracy the dynamics of the rotor angle $\theta_j$ and angular frequency $\omega_j$ of the generators/loads in real systems.

Lastly, we have included a noise term $\xi_j$ at each node.
In principle, we could ask why there is no noise term in the rotor-angle equation.
One can similarly include a noise element in the angles, which will manifest itself in some format on the frequency.
Yet here we arrive at the core of what stochastic noise $\xi_j$ is. 
On the one hand, we know we can broadly model complex power grids with simplified models as the swing equation~\eqref{eq:swing}.
On the other hand, thousands of power-generating units and millions of consumers are `condensed' into a single machine/load, a node in our network.
One manifest effect is that real-world recording of voltage angles, power-grid frequency, power flows, or any other physically relevant measures, are noisy.
These are not solely dynamic processes, these are stochastic processes, noisy processes, whence we can learn plenty from the system by examining the statistical properties of various real-world recordings~\cite{RydinGorjao2020b}.

\begin{figure*}[t]
    \includegraphics[width=\textwidth]{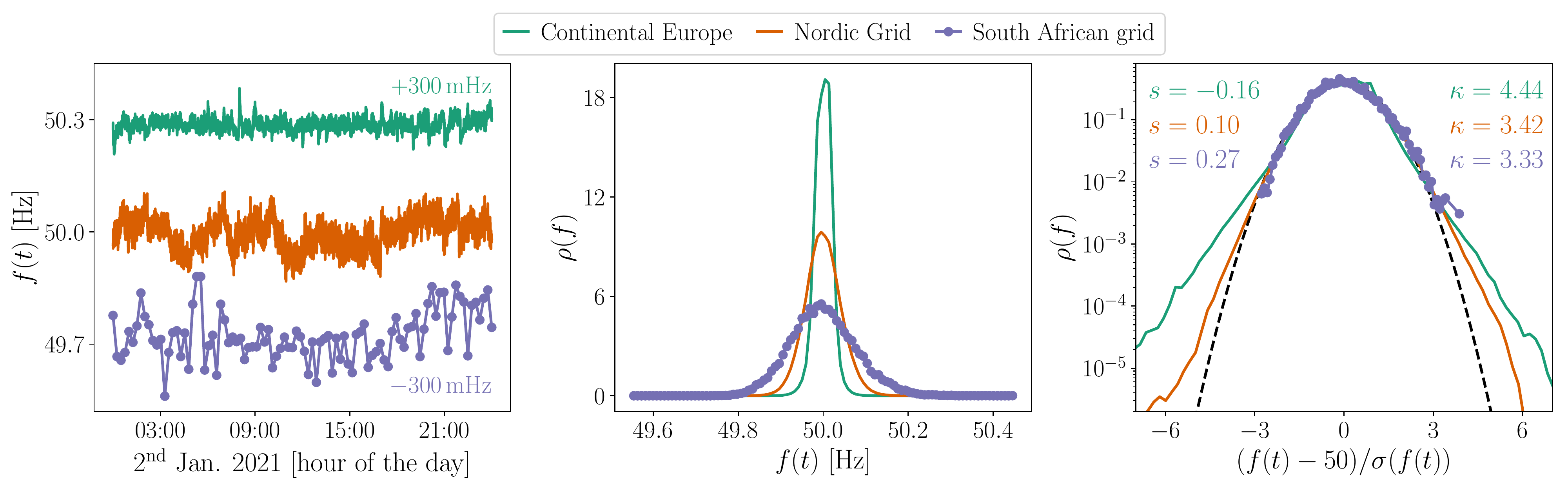}
    \caption{Three exemplary power-grid frequency trajectories (left) from Continental Europe (green), Nordic Grid (orange), and South African grid (purple), between the 2\textsuperscript{nd} and 3\textsuperscript{rd} of January 2021.
    The recordings are vertically displaced for clarity.
    The probability density functions for the whole of 2021 are shown in the centre and right panels, the latter in a vertical logarithmic scale. 
    All three power grids exhibit non-Gaussian phenomena, particularly showing larger than normal kurtosis ($\kappa>3$) and some skewness of the distributions.
    The power-grid frequency data from Continental Europe and the Nordic Grid have a temporal resolution (sampling rate) of $1$\,s.
    The data from the South African grid has $15$ minutes resolution ($900$\,s).
    All figures produced with \texttt{Matplotlib}, \texttt{NumPy}, \texttt{SciPy}, and \texttt{pandas}~\cite{Matplotlib, NumPy, SciPy, pandas}.}\label{fig:2}
\end{figure*}

\subsection{Fokker--Planck description of power-grid frequency recordings}
Returning to our noise-disturbed swing equation~\eqref{eq:swing} we can now focus on writing a Fokker--Planck equation for the probability density $\rho(\theta,\omega,t)$ for a single node~\cite{Acebron1998, Acebron2000, Rodrigues2016}.
Let us consider the noise is delta correlated $\langle \xi_i(t)\xi_j(t')\rangle = 2B\delta_{i,j}\delta(t-t')$, with $B$ the diffusion constant.
Now considering precisely the previous argument that our models are descriptive of $N\gg 1~~(N\to\infty)$, we can take the `mean-field' approximation yielding~\cite{Acebron1998, Acebron2000}
\begin{equation}
    \frac{\partial \rho}{\partial t} = -\omega \frac{\partial \rho}{\partial \theta} + \frac{1}{m}\frac{\partial }{\partial \omega}\left[\left(-d\omega + P_j^{\mathrm{m}} - P_j \right)\rho\right] + \frac{B}{m^2}\frac{\partial^2 }{\partial \omega^2}\rho.
\end{equation}
We assume that $M_j=m$ and $D_j=m$.
We can separate the probability density $\rho(\theta,\omega,t)=\phi(\theta,t)\eta(\omega,t)$ and focus solely on the statistics of the angular frequency $\eta(\omega,t)$, following Acebrón \& Spigler~\cite{Acebron1998}.
This results in
\begin{equation}\label{eq:fokker-planck_kuramoto}
    \frac{\partial \eta}{\partial t} = -\frac{d}{m} \frac{\partial}{\partial \omega}\omega\eta + \frac{B}{m^2}\frac{\partial^2 \eta}{\partial \omega^2}.
\end{equation}
The static solution at $t\to \infty$ (and having appropriate decay conditions of $\eta(\omega)=0$ at $\omega\to\pm\infty$) is the well known Gaussian distribution
\begin{equation}\label{eq:gaussian_kuramoto}
    \eta(\omega) = \sqrt{\frac{md}{2\pi B}}e^{-\frac{md}{2B}\omega^2}.
\end{equation}
There are various important details to notice here.
First of all, we have started with uncorrelated Gaussian white noise, which is the most natural assumption as it is the simplest noise model.
One can easily argue that power-grid frequencies are not Markovian processes and that these are likely driven by correlated noises~\cite{Friedrich2011, Tyloo2022a, SalcedoSanz2022, Kraljic2022}.
Non-Markovianity is ubiquitous across energy systems~\cite{RydinGorjao2021a, Han2022a, RydinGorjao2022}, yet we do not necessarily need to address it directly while modelling.
In this work, we are interested in understanding the statistics of power-grid frequency recordings in South Africa, for which we have access to recordings with a time sampling of $15$ minutes.
This time resolution is much beyond the intrinsic Einstein--Markov length of the process, i.e., the time scale at which a non-Markovian process can safely be modelled in a Markovian setting~\cite{Tabar2019}.
For an inhomogeneous spatially extended parabolic partial differential equation of coupled swing equations, see Ref.~\cite{Pagnier2022}.

\begin{figure*}[t]
    \includegraphics[width=\textwidth]{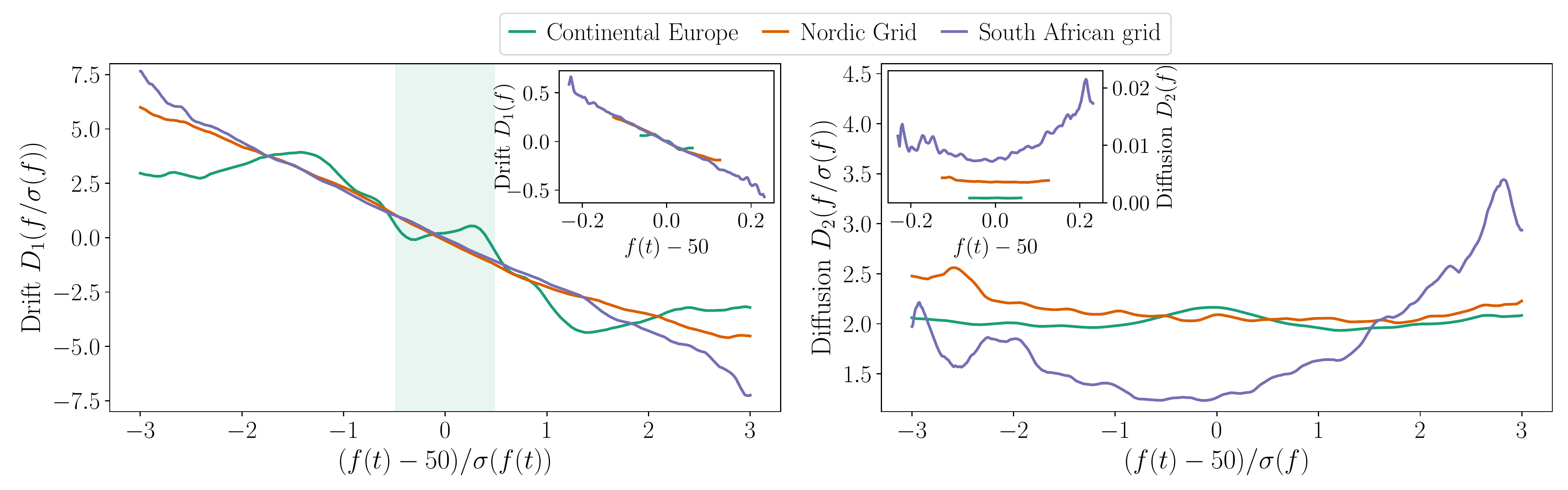}
    \caption{Drift and diffusion estimated for three power-grid frequency recordings from Continental Europe, the Nordic Grid, and the South African grid.
    (Left panel) The drift $D_1$ in the Nordic Grid and South Africa is linear with a negative slope, indicating a linear response, as expected in~\eqref{eq:fokker-planck_kuramoto}.
    The case of Continental Europe is less trivial, the shaded green area shows the area wherein there is no control active -- the `deadband'.
    (Right panel) the diffusion $D_2$ in Continental Europe and Nordic Grid can be said to be constant.
    In South Africa, we observe a functional dependence on angular frequency.
    Data covers the year 2021~\cite{RydinGorjao2019}.}\label{fig:3}
\end{figure*}

Secondly, we can already suspect that our Gaussian description of the distribution of the frequencies~\eqref{eq:gaussian_kuramoto} might not be sufficient.
Schäfer \textit{et al.}~\cite{Schaefer2018} report on more complex statistics of power-grid frequency in Continental Europe and the Nordic Grid, among others, and provide a descriptive analysis based on a convolution of statistics which results in power-grid frequency distribution far beyond Gaussianity.
Various subsequent works offer explanations for the aforementioned phenomena~\cite{RydinGorjao2020a, Farmer2021, RydinGorjao2021a}.
The most straightforward explanation of the non-Gaussianity of power-grid frequency statistics in Continental Europe and the Nordic Grid is due to changes in dispatch.
At fixed intervals, usually every hour, with some minor in-between blocks of 15 minutes, different power plants, big or small, sell their power generation capacities on an electricity exchange market.
This is a fairly complex problem of bidding and offering, which, for our purpose, boils down to one known and very clear effect: at every trading block (1 hour or 15 minutes), power-grid frequency sees a large overshoot or a large sag.
This effect comes about because the generation needs to adapt to an ever-changing consumption, that is, the generation adapts to the predicted consumption for the following 1-hour block.
If there is more generation than consumption, power-grid frequency goes up from its nominal value of $50$~Hz and vice versa, lower generation implies a sag of the frequency.

In Fig.~\ref{fig:2} (left) we display a snippet of the three frequency recordings Continental Europe, the Nordic Grid, and the South African grid, between the 2\textsuperscript{nd} and 3\textsuperscript{rd} of January 2021.
We can see that these recordings look different from one another.
In the centre panel, we display the probability density $\rho(f)$ of the three recordings, covering all of 2021.
Similarly, we exhibit the same thing in the right panel on a vertical logarithmic scale.
Here we can notice the evident heavy tails of all three recordings.
A purely Gaussian distribution is shown in black dashed lines and is an inverter parabola in a vertical logarithmic scale.

To best understand the distribution of power-grid frequency in the various grid, we can directly examine the skewness $s$ and kurtosis $\kappa$, given respectively by
\begin{equation}
\begin{aligned}
    s_X = \frac{\operatorname{E}\left[(X - \mu)^3\right]}{\left(\operatorname{E}\left[(X - \mu)^2\right]\right)^{3/2}}, \quad
    \kappa_X = \frac{\operatorname{E}\left[(X - \mu)^4\right]}{\left(\operatorname{E}\left[(X - \mu)^2\right]\right)^2}.
\end{aligned}
\end{equation}
A Gaussian distribution has skewness $s=0$ and a kurtosis $\kappa=3$.
That is, it is purely symmetric and does not have any heavy (or light) tails.
Being facetious, Gaussian distributions are very normal.
What we observe is that none of the recordings are symmetric ($s\neq 0$) and they all show heavy tails ($\kappa>3$).
The aforementioned changes in the dispatch/market activity in Continental Europe and the Nordic Grid explain this behaviour.
But what about South Africa?

\subsection{Understanding linear response and the effect of noise}

We have introduced a Fokker--Planck equation describing the probability density $\rho(\omega)$, wherein we can directly evaluate the effects of the inertia $m$ in the system and the presence of noise, mediated by its amplitude $B$.
We can now test, using our frequency recordings, how much our simplistic Fokker--Planck model~\eqref{eq:fokker-planck_kuramoto} can describe the data.
In a more general setting, we write a Fokker--Planck equation as~\cite{Risken1996, Gardiner2009, Tabar2019}
\begin{equation}\label{eq:fokker-planck}
    \frac{\partial \eta}{\partial t} =  \frac{\partial}{\partial \omega}D_1(\omega)\eta + \frac{1}{2}\frac{\partial^2}{\partial \omega^2} D_2(\omega)\eta,
\end{equation}
where $D_1(\omega)$ is known as the drift and $D_2(\omega)$ the diffusion.
In general, $D_n(\omega)$ are the Kramers--Moyal coefficients.
If we compare this with our Fokker--Planck equation, we see that
\begin{equation}
    D_1(\omega) = -\frac{d}{m}\omega, \quad D_2(\omega) = D_2 = \frac{B}{m^2}.
\end{equation}
First, we notice the linear response of the drift $D_1(\omega)$, which is a linear function of $\omega$.
Second, we note that given out choice of noise $\xi$ mediated by $B$, the diffusion $D_2(\omega) = D_2$ is not a function of the angular frequency.

We are now interested in obtaining the drift $D_1$ and diffusion $D_2$ directly from the frequency data.
We can make use of (Nadaraya--Watson) non-parametric estimators~\cite{Nadaraya1964, Watson1964} to extract the functional form of $D_1$ and $D_2$ from the data~\cite{Gottschall2008, Lamouroux2009, Anteneodo2009, Anteneodo2010, RydinGorjao2019, RydinGorjao2021b}.
For a time series $x(t)$, like power-grid frequency recordings, the Kramers--Moyal coefficients can be estimated using a kernel density estimation~\cite{Lamouroux2009}
\begin{equation}
\begin{aligned}
    D_m(x) &\sim \frac{1}{m!}\frac{1}{\Delta t}\langle (x(t+\Delta t) - x(t))^m|x(t) = x \rangle \\
    & \sim \frac{1}{m!}\frac{1}{\Delta t}\frac{1}{N}\sum_{i=1}^{N-1} K(x - x_i)(x_{i+1}-x_{i})^m,
\end{aligned}
\end{equation}
where $K(x)$ is a normalised kernel with a given bandwidth $h$ (similar to what the number of bins is in a histogram), such that $K(x) = 1/h K(x/h)$.
We use the commonly employed Epanechnikov kernel $K(x) = 3/4 (1-x^2)$, with support $|x|<1$~\cite{Epanechnikov1967}.
Furthermore, we apply some finite-time corrections to improve the quality of the estimations by accounting for the fact that the recordings have a finite-time sampling (cf. Refs.~\cite{Gottschall2008, RydinGorjao2021b}).

In Fig.~\ref{fig:3} we display the drift $D_1(f/\sigma(f))$ over the normalised power-grid frequency $f$ (left) and the diffusion $D_2(f/\sigma(f))$, also normalised (right).
In the insets, the non-normalised versions of each aforementioned Kramers--Moyal coefficient can be found.

Let us start by pointing out what immediately agrees with our understanding of power-grid frequency.
Both the Nordic Grid and the South African grid show a clear linear response of the drift.
This response almost overlaps, which points to an identical ratio of the damping and inertia $d/m$, which is also expected, as these go hand-in-hand.
When we examine the Continental European grid, we find a much stranger picture.
Continental Europe has a band of frequency where no control is active, i.e., there is no active control to maintain the frequency at its nominal value, marked in the shaded area on the left panel, known as the `deadband'~\cite{Vorobev2019}.
In the drift we even seem to see and effect of `repulsion', i.e., the drift has a positive slope, pushing the frequency off from the nominal value.
Moreover, at frequency deviations larger than $|\sigma(f)|>1.5$, it also seems like there is no control or linear response.
This agrees to some extent with the control mechanism in Continental Europe denoted as `secondary control', which is a slow mechanism to restore large frequency deviations to a nominal value.
Being a slow mechanism, it cannot be captured in our stochastic approach and functions as a slow dynamical effect.
We should note that the Nordic Grid and South African systems similarly have their own `deadband', as well as primary and secondary control mechanisms, yet these are not detectable using our analysis.

\begin{figure*}[t]
    \includegraphics[width=\textwidth]{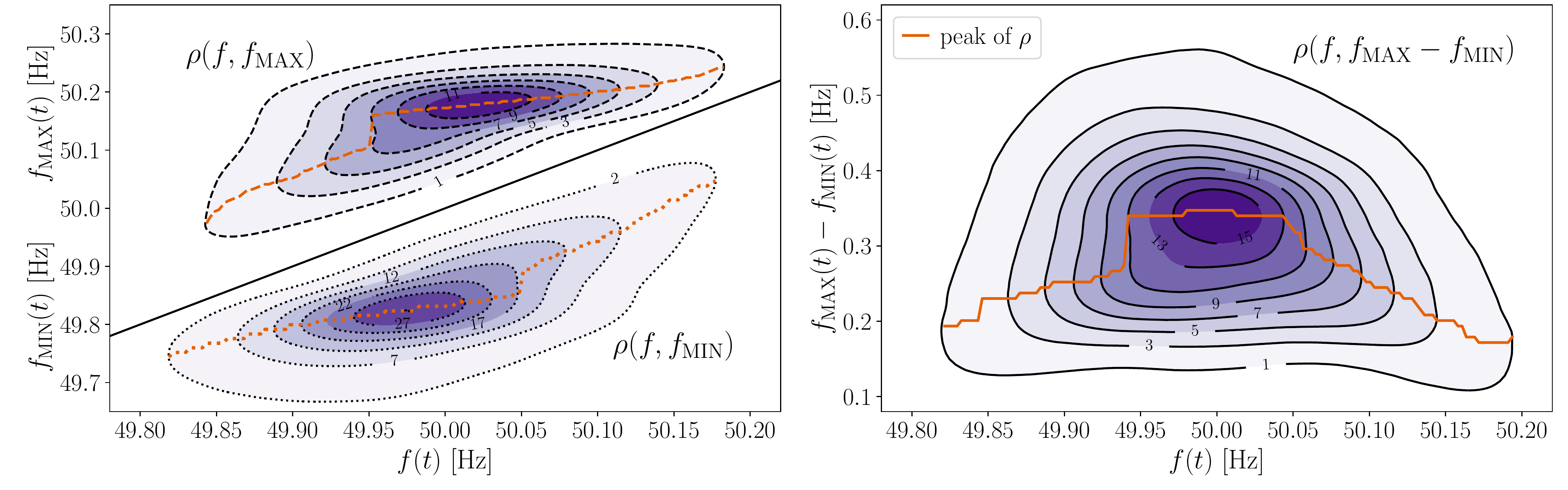}
    \caption{Probability of occurrence of sags or overshoots in relation to the recorded power-grid frequency in South Africa in 2021.
    (Left panel) Two-dimensional probability density of the recorded frequency $f(t)$ and the minimal $f_{\mathrm{MIN}}$ and maximal $f_{\mathrm{MAX}}$ recorded frequency.
    (Right panel) Two-dimensional probability density of the recorded frequency $f(t)$ and the difference $f_{\mathrm{MIN}} - f_{\mathrm{MAX}}$.
    When the frequency is far from the nominal value of $50$\,Hz, the minimal and maximal frequencies come together, pointing to a strong control at play when the system is far from equilibrium.}
    \label{fig:4}
\end{figure*}

When we examine the diffusion $D_2(\omega)$ of the three recordings, seen on the right panel of Fig.~\ref{fig:3}, we observe an almost constant value for both Continental Europe and the Nordic Grid.
The case of the South African grid is a bit different.
The diffusion is state-dependent, i.e., the value of $D_2(\omega)$ is roughly proportional to $\omega^2$ and not a constant.
Thus, at first glance, it seems that when the South African power-grid frequency is not at its nominal value of $50$~Hz but either slightly above or slightly below, the amplitude of the noise is larger.
We see this contrasts with our initial simplistic assumption when writing~\eqref{eq:fokker-planck_kuramoto}, wherein we started with simple \textit{state-independent} or \textit{additive} Gaussian white noise.
This does not seem to be the case in South Africa.
Thus, we return to~\eqref{eq:fokker-planck} and write a more general static solution to the Fokker--Planck equation as
\begin{equation}\label{eq:gaussian}
    \eta_{D_{2}}(\omega) \approx \sqrt{\frac{d}{2\pi m D_2(\omega)}}e^{-\frac{d}{2 mD_2(\omega)}\omega^2},
\end{equation}
wherein we have that the variance of $\eta(\omega)$ depends on the angular frequency.
This is the static solution to our swing equation with noise~\eqref{eq:swing} with \textit{multiplicative} noise (or state-dependent noise).
We note that this is only approximate since this density is not always normalisable.
In fact, with $D_1(\omega)$ a linear function of $\omega$, the underlying equation of motion is identical to an Ornstein--Uhlenbeck process, from which we know its closed-form solutions~\cite{Risken1996, Gardiner2009, Tabar2019}.
Unfortunately for the simple case of multiplicative/state-dependent noise $D_2(\omega) = b+c\omega^2$, with $b, c>0$ two constants, that resemble our case in South Africa, one cannot evaluate the expected values $\operatorname{E}[X^n]$ explicitely.
We can nevertheless approximate the density function for small $c>0$.
Let us consider the infinite Taylor expansion of $\eta(\omega)$ for small $c$, denoted $\eta_c(\omega)$, given by
\begin{equation}
\begin{aligned}
    \eta_c(\omega) &= \eta(\omega)\left[1 + \frac{c}{2b}\omega^2\left(\frac{d\omega^2}{mb} -1\right)  \right.\\
    &\qquad\qquad\quad +\left. \frac{1}{2}\frac{c^2}{4b^2}\omega^4\left(\frac{d^2\omega^4}{m^2b^2} - 6\frac{d\omega^2}{mb} + 3\right) + \dots\right] \\
    & = \eta(\omega)\sum\limits_{n=0}^{\infty}\frac{z^{2n}}{n!}\operatorname{He}_{2n}(z) \left(\!\sqrt{\frac{mc}{2d}}\right)^{2n}\!\!\!\!\!\!, ~~\mathrm{with}~z^2=\frac{d\omega^2}{mb},
\end{aligned}
\end{equation}
with $\operatorname{He}_{2n}(z)$ the Hermite polynomials.
From the generating function of the Hermite polynomials
\begin{equation}
    e^{\frac{1}{2}zt - t^2} = \sum\limits_{n=0}^{\infty}\frac{\operatorname{He}_{n}(z)}{n!} t^{n}
\end{equation}
we get
\begin{equation}
    e^{-\frac{t^2}{2}}\cosh(zt) = \sum\limits_{n=0}^{\infty}\frac{\operatorname{He}_{2n}(z)}{(2n)!} t^{2n},
\end{equation}
from which we draw our approximate expression by substituting $t=\sqrt{\frac{mc}{2d}}z=\sqrt{\frac{c}{2b}}\omega$
\begin{equation}
    \eta_c(\omega) = \eta(\omega)e^{-\frac{c\omega^2}{4b}}\cosh\left(\frac{\sqrt{dc}}{\sqrt{2m}b}\omega^2\right).
\end{equation}
With this closed-form approximation of the probability density $\eta(\omega)$ for small $c$ we can now find the expected values $\operatorname{E}[X^n]$, from which we obtain a kurtosis $\kappa_X^c$
\begin{equation}
    \kappa_X^c = 3.2140,
\end{equation}
with $d/m=0.6723$, $b=0.0023$, and $c=0.0467$ estimated from the drift and diffusion.
This is close to our empirical value of $\kappa=3.33$.
We should note that this is a crude approximation, but it helps understand the effect of multiplicative noise on generating leptokurtic distributions (i.e., distributions with kurtosis $\kappa>3$).

We will now contrast our analysis with the aid of two other power-grid frequency recordings from South Africa.

\subsection{Counter-intuitive frequency deviations}

We now address an interesting -- and quite a counter-intuitive effect -- in power-grid frequency recordings in South Africa.
Alongside recordings of the frequency in South Africa with a sampling of $15$ minutes, we have access to the smallest and largest recorded frequency measurements within each $15$-minute block.
Therein we have three time series we can examine:
The frequency $f(t)$ we have thus far examined, the recorded maximal frequency $f_\mathrm{MAX}(t)$, and the recorded minimal frequency $f_\mathrm{MIN}(t)$.

In Fig.~\ref{fig:4} we display the two-dimensional probability density functions of $\rho(f,f_\mathrm{MAX})$ and $\rho(f,f_\mathrm{MIN})$ (left) and the difference $f_\mathrm{MAX}-f_\mathrm{MIN}$ (right).
If we focus on the left panel, we can see that an increase of the frequency $f$ similarly increases both the $f_\mathrm{MIN}$ and the $f_\mathrm{MAX}$.
This is somewhat counter-intuitive.
If the frequency $f$ is \textit{above} its nominal value of $50$~Hz within a period of 15 minutes we expect that the maximal frequency $f_\mathrm{MAX}$ follows this effect, but not necessarily the minimal frequency $f_\mathrm{MIN}$.
In the end, one would hope that the frequency is brought down back to $50$\,Hz in the span of $15$ minutes.
This follows in reverse if the frequency drops below $50$\,Hz.

One potential explanation for this would assume that we do not have a stationary process, but instead that the average (on the centre of the distribution) is moving with the frequency.
This supports the idea that what large overshoots and nadirs of the frequency we observe in the South African grid follow a similar phenomenon to those in Continental Europe and the Nordic grid: These are dynamical effects of the market.
This being the case, we would expect that the minimal and maximal frequency would shift in tandem.
If there is too much generation in the system the minimal frequency is likely to also have increased.
Similarly, too much consumption brings down the maximal frequency.

What remains to be explained is shown in the right panel of Fig.~\ref{fig:4}.
Here we see the difference between $f_\mathrm{MAX}$ and $f_\mathrm{MIN}$.
We see that when the system is out of equilibrium these two measured frequencies come together.
This points to a strong control mechanism starting up as the frequency deviates from its nominal value.
Yet this is in clear contradiction with our previous result wherein we showed there is multiplicative noise in the system.
If indeed there is multiplicative noise at frequencies far from nominal the distribution must widen and, statistically speaking, $f_\mathrm{MAX}$ and $f_\mathrm{MIN}$ should grow further apart, not come together.

At the present state, we cannot offer an explanation for this.
The obvious remark points to the low time sampling of the frequency data, which inevitably leads to finite-time effects when estimating the drift and diffusion~\cite{Gottschall2008, Anteneodo2009, RydinGorjao2021b}.
Possibly at higher sampling rates of the power-grid frequency this effect subsides.

Altogether, we have offered a first explanation for the non-Gaussian distribution of the power-grid frequency recordings in South Africa -- with its various caveats.
Yet, as we have seen for Continental Europe and the Nordic Grid, noise in power-grid frequency does not seem to be multiplicative/state-dependent, but constant across the frequencies.
Moreover, we have also discussed that these recordings similarly have high kurtosis and that this arises primarily due to changes in dispatch and market effects.
In South Africa, similar large deviations of power-grid frequency -- which lead to high kurtosis of the distribution -- can occur due to scheduled and unscheduled load-shedding and rolling blackout events.
We now explore this avenue of effects on a coarse-grain modular power-grid model of South Africa with numerical solvers of the swing equation with noise~\eqref{eq:swing}.

\subsection{The effects of rolling blackouts and load shedding}\label{sec:num_sim}

\begin{figure}[t]
    \includegraphics[width=\columnwidth]{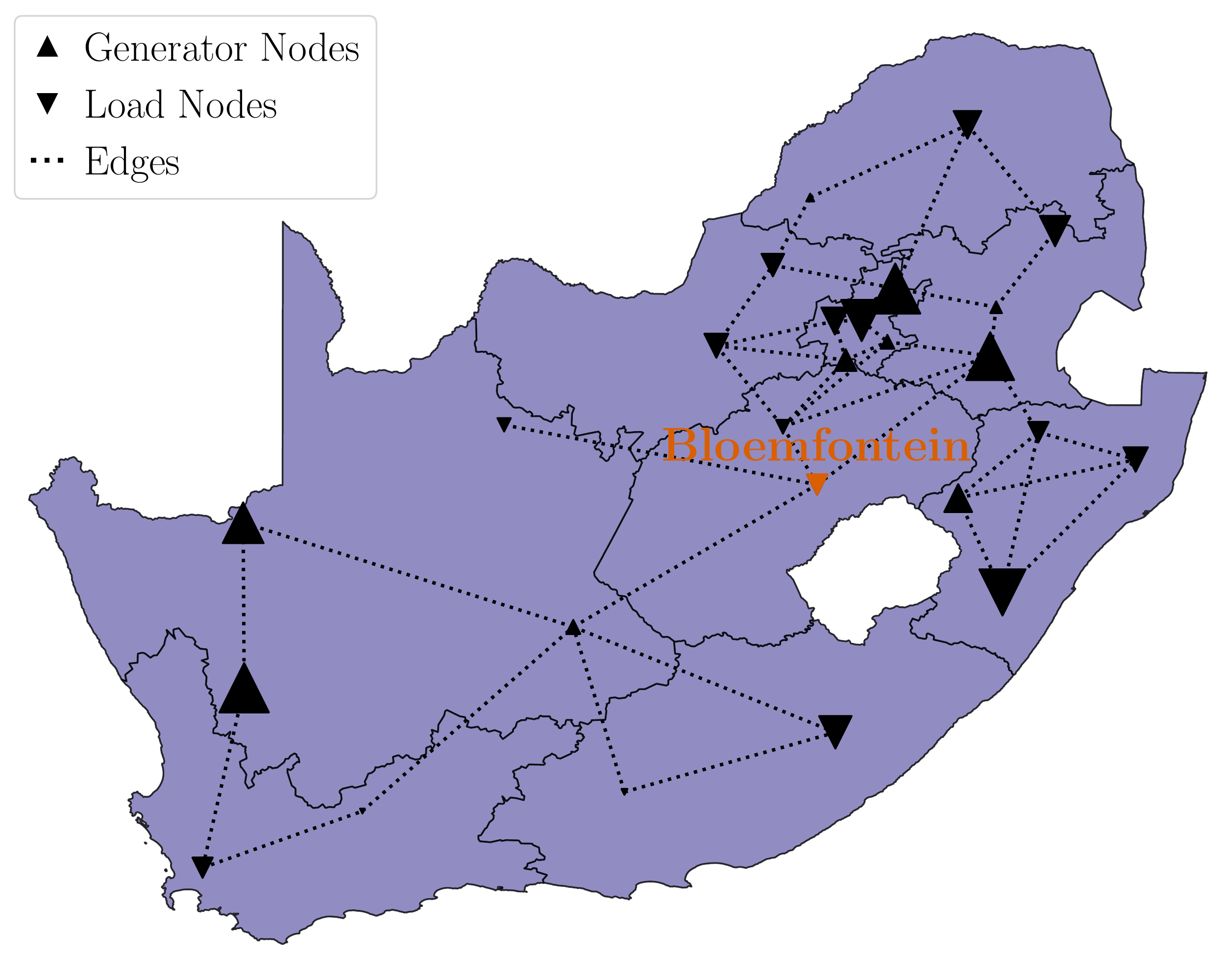}
    \caption{Model power-grid network of the South Africa Grid based on \texttt{PyPSA-ZA}~\cite{Hoersch2017}, with 26 nodes and 41 lines. 
    The classification into generator and load nodes is based on various available data~\cite{Hoersch2017, WikipediaPowerPlantsSA, WorldPop} and a power-flow balancing obtain with \texttt{pandapower}~\cite{pandapower2018}.}
    \label{fig:5}
\end{figure}

In order to simulate power-grid voltage frequency dynamics, we make use of a coarse-grain model of the South African grid introduced by Hörsch \& Calitz~\cite{Hoersch2017}.
The model comprises $26$ geographically distributed nodes, located roughly at the cultural centres of the $9$ provinces of South Africa, coupled via $41$ aggregated high-voltage transmission lines.
A representation of the model is given in Fig.~\ref{fig:5}.
Each node in this power-grid model is considered as modelled via the swing equation~\eqref{eq:swing} diffusively coupled following~\eqref{eq:power_exchange}.
The total capacity at each node is assessed via the freely available data from the lists of power plants in South Africa found in Wikipedia~\cite{WikipediaPowerPlantsSA}.
The capacity of each power plant is added to the nearest geographic node of the power-grid model.
To account for the consumption, the estimation of the population in 2021 in each region nearest to a node is also summed~\cite{WorldPop}.
The total electricity demand for 2021 was $219\,423$~GWh, following ESKOM's 2021 Integrated Report~\cite{ESKOMReport2021}, from which we estimate the total average power consumption and split it according to each node's population.
Inertia $M_i$ and damping $D_i$ for each node $i$ is estimated following Pagnier \& Jacquod~\cite{Pagnier2019}.
The transmission line parameters, i.e., the susceptance, are given in Ref.~\cite{Hoersch2017}.

To simulate the frequency (and rotor-angle) dynamics we first solve the alternating current (AC) power-flow problem using \texttt{pandapower}~\cite{pandapower2018}.
It, firstly, ensures the physical power-flow problem is solvable, but also adequately balances generation and consumption at each node, resulting in the generated power $P_j^{\mathrm{m}}$ at each node.
Nodes with negative generation, i.e., consumption, act as loads in the system.
The sketch of which nodes act as generator and which act and load is given in Fig.~\ref{fig:5}.
For the numerical simulations we assume a lossless system, i.e., $G_{j,\ell}=0$, a fairly common assumption in high-voltage transmission grids.
We consider as well a small Gaussian white noise added to each node to account for some stochasticity in the problem.

The target is now to model load-shedding events to showcase how these can influence power-grid frequency dynamics.
As we have little access to the actual operation, implementation, phasing, and duration of the load-shedding events, we focus instead on matching the few statistics we do have access: the number of hours under different load-shedding stages in South Africa.
For reference, we showcase in Tab.~\ref{tab:1} the years 2019, 2020, and 2021.

\begin{table}[t]
    \centering
    \begin{tabular}{|c|c|c|c|c|c|c|c|}
         \hline
         year & \thead[c]{St. 1 \\\relax [GWh]} & \thead[c]{St. 2 \\\relax [GWh]} & \thead[c]{St. 3 \\\relax [GWh]} & \thead[c]{St. 4 \\\relax [GWh]} & \thead[c]{St. 5 \\\relax [GWh]} & \thead[c]{St. 6 \\\relax [GWh]} & \thead[c]{\# \\\relax hours} \\ \hline
         2019 & 43 &  618 &  93 & 568 & -- & 30 & \textit{530} \\
         2020 & 133 & 1192 & 141 & 332 & -- & -- & \textit{859} \\
         2021 &  \textbf{79} & \textbf{1848} & \textbf{210} & \textbf{384} & -- & -- & \textbf{\textit{1169}} \\ \hline
    \end{tabular}
    \caption{Total estimated load shed energy as reported by the \textit{Council for Scientific and Industrial Research -- Energy Centre} of South Africa (see p. 171 in Ref.~\cite{SAloadshed}) at different load-shedding stages (St. 1 to St. 6). 
    Shown as well are the total number of hours under load-shedding in South Africa in 2019, 2020, and 2021.}
    \label{tab:1}
\end{table}

\begin{figure}[t]
    \includegraphics[width=\columnwidth]{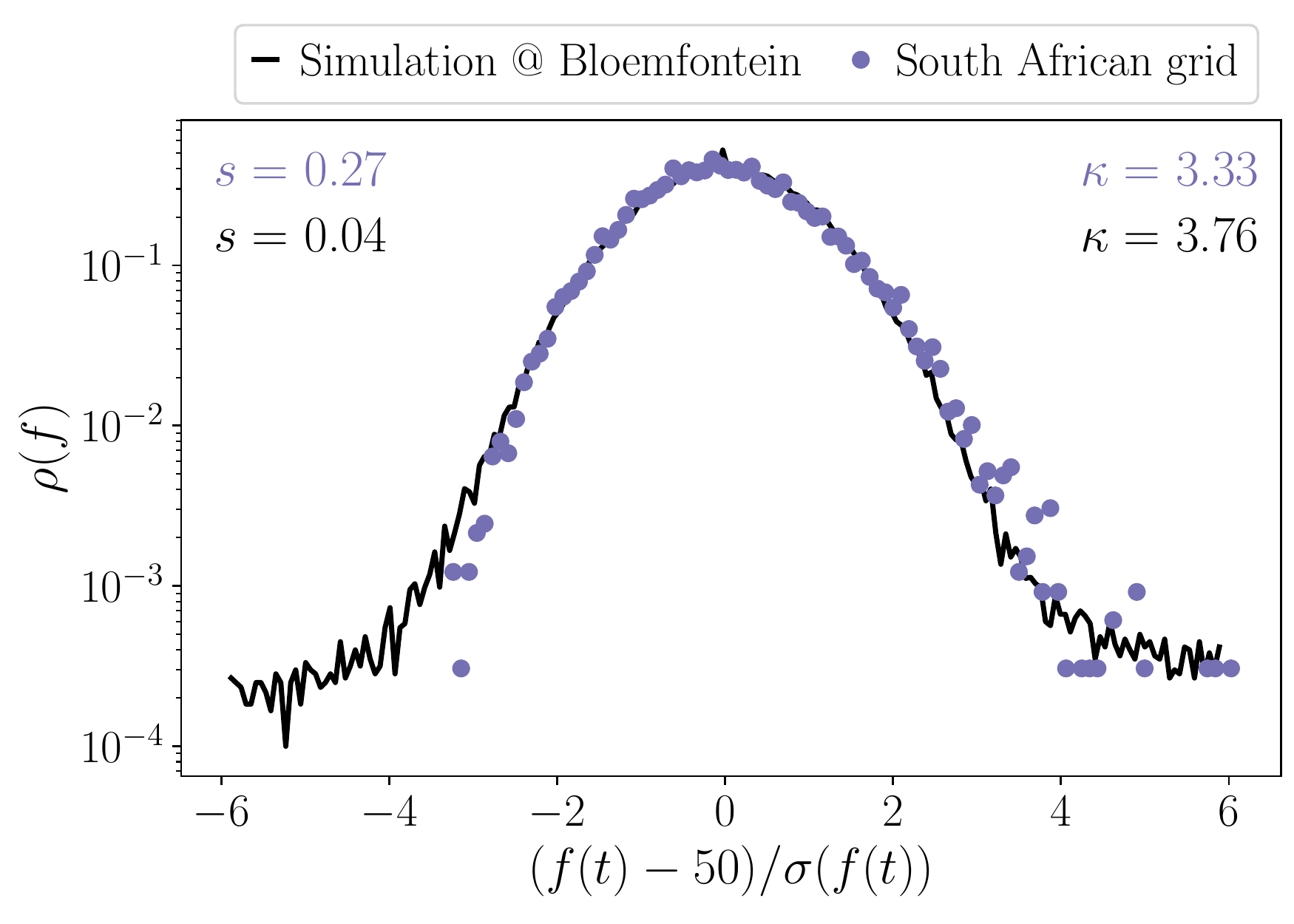}
    \caption{Numerical simulation of power-grid frequency obeying the swing-equation model~\eqref{eq:swing} with 26 machine/nodes and 41 transmission lines/edges. 
    The simulated system is affected by load-shedding events 39 out of 283 hours ($\sim13.8\%$ of the time).
    We can observe a direct effect on the kurtosis $\kappa$ as large excursions of the frequency take place as the power flow readjust to a rapid change of the load.
    The data is truncated at $\pm~6$ standard deviations as various control mechanisms would enter into frequency control that cannot be easily simulated with~\eqref{eq:swing}.}\label{fig:6}
\end{figure}

To model the load-shedding events, we consider an abrupt change in demand by randomly selecting a node and a random subset of its neighbours.
The total load-shedding is split uniformly across these nodes and the power-flow balance is recalculated.
To mimic the system close to reality, we simulate a similar ratio of load-shedding events covering $\sim\!\!13\%$ of time in a simulation lasting 336 hours.
The change in power leads to large frequency excursions -- one of the causes for the elevated kurtosis of the frequency statistics.
In Fig.~\ref{fig:6} we display the statistics of the modelled system as measured at the node located in Bloemfontein (see Fig.~\ref{fig:5}).
We can observe a direct impact on the kurtosis as the model system adjusts to the change in the power flows due to the sudden events of load shedding.
This leads to large frequency excursions and thus elevated kurtosis.
We note that in our simulations the system never loses power balance, simply readjusted generation directly as demand drops.
The dynamical simulations simply readjust the angle and frequency dynamic variables to these changes.

One can easily argue that most likely load shedding leads to a temporarily unbalanced system -- which would imply the dynamical system would attain a new fixed point not at $f=50\,$Hz (or $\omega=0$).
This, arguably, would explain the data better -- yet numerically simulating unbalanced systems results in the loss of stable fixed points and the uncoupling of some of the nodes in the system.
Thus, we pursue a simpler approach and consider an abrupt change of flows, which inherently maintain power balance (i.e., the system converges back to $\omega_i(t)=0$ after some time) but affects the angles $\theta_i$, leading to excursions of the angular frequency.
Overall, as expected, the statistics of the frequency show an increased kurtosis $\kappa_{\mathrm{ZA}}^{\mathrm{sim}}=3.76$, which, in our limited simulations, presents itself above the empirical kurtosis $\kappa_{\mathrm{ZA}}=3.33$.
One of the main reasons for such an elevated kurtosis in the numerical simulation is that this leads to large angular frequency excursions in both positive and negative directions.

\section{Conclusion}
In this work we have examined power-grid frequency recordings from the South African grid for the year of 2021, having a 15-minute temporal resolution.
We contrasted these recordings with similar recordings from Continental Europe and the Nordic Grid, both having a much higher temporal resolution of 1 second.
At this much coarser time resolution, we observe very similar statistics of power-grid frequency in South Africa as in Europe.
All distributions have heavy tails (kurtosis $\kappa>3$) and are slightly skewed, with South Africa having a kurtosis of $\kappa_{\mathrm{ZA}}=3.33$.

Using a simplistic Fokker--Planck equation to describe the statistics of power-grid frequency, we examine the drift and diffusion of all three recordings via a non-parametric estimation of the Kramers--Moyal coefficients.
Overall, we find an identical response mechanism in the recordings, which is given by the ratio of overall damping and inertia in the system.
That is to say, for all recordings we estimate an identical linear-restoring-force mechanics of the frequency pushing the system to the nominal value of $50$\,Hz.
Examining the diffusion in the recordings we come across what resembles multiplicative/state-dependent noise in the South African recordings.
We derive an approximate analytical expression to treat multiplicative noise and show that with such an expression one can estimate the statistical moments $\operatorname{E}[X^n]$.
From this expression, we obtain our first explanation for the large kurtosis, resulting in $\kappa_{\mathrm{ZA}}^c=3.21$.

This, however, disagrees with a subsequent analysis of two additional South Africa power-grid time series we include in this work.
Having access to both the minimal and maximal measure power-grid frequency in each 15-minute block of 2021, we find a counter-intuitive relation of their statistics when the frequency is far from its nominal value.
Away from equilibrium, the minimal and maximal power-grid frequency recordings become closer to each other -- this disagree with our Fokker--Plank approach and contrasts with the presence of multiplicative noise, which should drive the maximal and minimal frequency apart, not together.

The narrowing of the minimal and maximal frequency, along with other details, points to a simpler explanation for the non-Gaussian statistics of power-grid frequency in South Africa:
The detrimental effect of unscheduled load shedding.
Interestingly, large frequency deviations in Europe are explained by various dispatch changes and electricity market activities (cf. Ref.~\cite{Schaefer2018,RydinGorjao2020a}), which, from a point-of-view solely of power-grid frequency, means large deviations are induced by strong changes in generation and consumption -- much alike load shedding events.
To this effect, we simulate a model power-grid system of South Africa composed of 26 nodes and 41 transmission lines.
We consider a losses system of inertial masses coupled in the network obeying a swing equation, wherein generation and consumption are drawn from freely available data of the power grid and the population is South Africa.
In order to mimic the effects of load-shedding we simulate the system as affected by drops in the load for roughly $13\%$ of the time.
We consider the effects at a central location in South Africa and show that the simulation lead as well to an increased kurtosis of $\kappa_{\mathrm{ZA}}^{\mathrm{sim}}=3.76$.

Overall we provide a explanation for the statistics of power-grid frequency in South Africa -- which shows similar characteristics to more studied power grids in Europe -- yet this grid has contended (and still contends) with various load-shedding events, which invariantly affects frequency stability.
We provide two explanations for the presence of elevated kurtosis of the frequency statistics in South Africa.
We hope this work can bring into focus hitherto overlooked power grids around the world, particularly those that have been or are afflicted by serious infrastructural, economical, or societal problems.

\begin{acknowledgements}
We thank Pedro G. Lind and Philipp C. Böttcher for their early suggestions and remarks on the manuscript.
\end{acknowledgements}

\bibliography{bib}

\end{document}